\documentclass[pre,twocolumn,showpacs]{revtex4}
\usepackage{epsfig}
\begin{document}
\title{
Theory of Single Photon Control from a Two Level System Source
}
\author{Yong He$^{1}$, Eli Barkai$^{1,2}$}
\affiliation{\ $^{1}$ Department of Chemistry and Biochemistry, Notre Dame University, Notre Dame, IN 46556 \\
\ $^{2}$ Department of Physics, Bar Ilan University, Ramat Gan 52900, Israel}
\begin{abstract}

 Generation of a single photon or a pair of photons from  a single
emitter is important for quantum information  applications.
Using the generating function formalism we investigate 
the theory of a few photons on demand for the square laser
pulse and the rapid adiabatic following method. 
Exact theory and numerical solutions  are used
to design control fields for a two level emitter,
which yield an  
optimal single or two photon source, 
under the constrains of finite laser field strength 
and finite interaction time. 
Comparison to experiments of Brunel et al, shows that 
the experiments were made close
to optimal conditions.  

\end{abstract}

\pacs{42.50.-p, 33.80.-b,32.50.+d}


\maketitle

 The generation of a single photon on demand, is motivated
by fundamnetal interest in quantum properies of light \cite{Wolf}, as well
as possible applications like quantum computation, cryptology, and
communication \cite{Moerner,Lounis}. 
Single photon experiments  using two level 
rubidium atoms \cite{Darquie}, single organic 
molecules \cite{Martini,Brunel,Lounis1}, 
and single quantum dots \cite{Michler}, 
are nowadays considered  in detail
due to these applications. 
Besides single photon sources, sources of pairs of photons are also
important, due to useful properties
of indistiguishable photons and entanglement \cite{HM,Santori1,Brattke}.
A generic problem in
single  photon
sources, are fluctuations in the number of photons emitted, and in
the time of emission, due to the quantum nature of the emission
process.

Let $P_N(t)$ be the probability of $N$ emission events in the time
interval $(0,t)$.
The generating function 
$ 2 { \cal Y } (s,t) \equiv
 \sum_{N=0}^\infty s^N P_N(t)$ contains information on photon
statistics necessary for the determination of $P_N(t)$ \cite{remark}.
For a two level atom or molecule, within the rotating
wave approaximation, the generalized optical Bloch equations
which yield the generating function ${\cal Y}(s, t)$ are
\cite{Cook,Brown,Brown1,Mukamel}
\begin{equation}
\begin{array}{l}
 \dot{{ \cal U}}\left(s,t\right) = - {\Gamma \over 2} {\cal U}\left(s,t\right)
+ \delta(t) {\cal V}\left( s,t \right) \\
\ \\
 \dot{{\cal V}}\left(s,t\right) = -
\delta(t) {\cal U}\left( s,t \right)-{\Gamma\over 2} {\cal V}\left( s,t \right)
- \Omega(t) {\cal W} \left( s,t \right) \\
\ \\
 \dot{{\cal W}}\left(s,t\right) =
\Omega(t) {\cal V}\left( s,t \right)-{\Gamma\over 2} \left( 1 + s \right)
 {\cal W}\left( s,t \right)
- {\Gamma \over 2} \left( 1 + s\right) {\cal Y} \left( s,t \right) \\
\ \\
\dot{{\cal Y} } \left( s,t \right) = - { \Gamma \over 2} \left( 1 - s \right)
{\cal W} \left( s,t \right) - { \Gamma \over 2} \left( 1 - s \right) {\cal Y} \left( s,t \right),
\end{array}
\label{eqA01}
\end{equation}
where $\Gamma$ is the spontaneous emission rate, and
$\Omega(t),\delta(t)$ are time dependent Rabi frequency and detuning.
One goal of this paper 
is to use Eq. (\ref{eqA01}) and find
optimal control fields $\delta(t)$ and $\Omega(t)$  which  
under physical constrains e.g. not too strong
laser fields, yield a few photons. 
Interaction of the molecule with its environment, for example
effects of spectral diffusion \cite{Brown,YongPRL}, 
and other sources of stochasticity 
\cite{BarkaiRev}  like triplet blinking,
are not considered here.
Still as we show later the
generating function formalism is in excellent agreement with
cryogenic temperature single molecule experiments.
 
 Two types of control approaches are used so far in experiments.
The first is the rapid adiabatic following method \cite{Brunel}, in this method
the single molecule is interacting with a continuous
wave (cw) laser field, and at the same time the absorption
frequency of the molecule in modulated periodically using a slow
radio frequency Stark field. 
In this case the Rabi frequency 
$\Omega(t)=\Omega$ is time independent,
and  
$\delta(t) = {\Delta_{\rm RF} \over 2} \cos\left( \nu_{\rm RF} t \right)$ 
\cite{Brunel}.
The
time dependent detuning $\delta(t)$ is designed to
bring the molecule in and out
of resonance with the cw laser in such a way that
a few photons are emitted per crossing (see details below). 
In a second very common approach the emitter interacts with a sequence
of laser pulses. 
We model such behavior by 
the text book example of a square pulse where the Rabi
frequency is 
\begin{equation}
\Omega(t) =\left\{
\begin{array}{l l}
\Omega & \ \  0 < t< T \\
0 &  \ \  T< t,
\end{array}
\right.
\label{eqS01}
\end{equation}
and the detuning $\delta(t)=0$. 
The specific goal  in applications varies, 
some experiments
are interested in single photon sources, while others consider two photon
sources.

The goal of theory of single photon control is to find 
the probability of emission of $N=0,1,2$ photons.
And then to design the optimal  external fields $\delta(t),\Omega(t)$
to maximize
the probability of single photon emission or the probability
of the emission
of a pair photons. Since very little is known on this
important control problem, we consider here simple cases.  
In particular,  the optimization of $P_1$ and $P_2$ 
under the constrain of finite time and strength of the excitation
laser field is investigated.  

%

We first investigate the square pulse and later consider
the rapid adiabatic following method. 
Only the limit of
long measurement time $t >> T$ is considered, for the sake of space. 
The details on the calculation of $P_N$ will be published elsewhere,
this calculation  becomes cumbersome already for small $N$.
We use dimensionless units with  $\Gamma=1$,
the notation $y =\sqrt{1 - 4 \Omega^2}$
and $x=\sqrt{1 - 16 \Omega^2}$, and find

\begin{equation}
P_0 = {e^{ - {T \over 2} } \over 2 y^2} \left[
\left(1 + y^2 \right)\cosh( \frac{T\,y}{2}) + y^2 - 1 + 2 y 
\sinh( \frac{T\,y}{2}) \right],
\label{eqMR0}
\end{equation}

\begin{widetext}
\begin{equation}
 P_1=  
  \frac{ \left( 1 - y^2 \right)e^{-\frac{T}{2}}} 
{8\,      y^5}
       \left[ y\,\left( -8 + T + 4\,y^2 + T\,y^2 \right) \,
          \cosh (\frac{T\,y}{2}) +
         2\, y\,\left( 4 + T - 2\,y^2 - T\,y^2 \right)  + 
            2\, \left( -3 + \left( 1 + T \right) \,y^2 \right) \,
             \sinh (\frac{T\,y}{2})  \right], 
\label{eqMR1}
\end{equation}



$$ P_2  = \frac{\left(y^2-1\right)^2 e^{-\frac{T}{2}}}{64 y^8}
 \left\{ \left[T (T+8) y^4+\left(T^2-28
    T-32\right) y^2+96\right] \cosh \left(\frac{T y}{2}\right)+ \right. $$

\begin{equation}
\left. 4
    \left[T \left(T+4\right) y^4-\left(T^2+8 T-8\right) y^2-24\right]+
2 y \left[\left(T^2 - 24 \right) y^2 
    -T \left(y^2+9\right)+60\right] \sinh \left(\frac{T
    y}{2}\right)\right\},
\label{eqMR2}
\end{equation}

\begin{equation}
\langle N \rangle = { \Omega^2 \over \left( 1 + 2 \Omega^2\right)^2} 
\left\{ T \left( 1 + 2 \Omega^2\right) - 2 + 2 \Omega^2 + e^{ - 3 T \over 4} \left[
\left( 2 - 2 \Omega^2\right) \cosh \left( \sqrt{ 1 - 16 \Omega^2} { T \over 4} \right)+ \left( 2 - 14\Omega^2 \right) { \sinh\left( \sqrt{ 1 - 16 \Omega^2} { T \over 4} \right) \over \sqrt{1 - 16 \Omega^2} } \right] \right\}, 
\label{eqAVN}
\end{equation}

$$ \langle N(N-1) \rangle= { ( 1 - x^2)^2 \over 8} \left\{ e^{ - { 3 T \over 4} +{ x T \over 4} } { 12 - \left( 16 + 3 T \right) x - 2 \left( 6 + T \right) x^2 + T x^3 \over \left( - 3 + x \right)^4 x^3 } +  \right. $$ 
\begin{equation}
\left.
e^{ - { 3 T \over 4} - { T x \over 4} } { - 12 - (16 + 3 T) x + 2 ( 6 + T) x^2 + T x^3 \over \left( 3 + x\right)^4 x^3 } 
+2 { \left[ T^2 ( - 9 + x^2)^2 + 32 ( 63 + 5 x^2) + 2 T ( - 351 + 30 x^2 + x^4) \right] \over (- 9 + x^2)^4} \right\}.
\label{eqnnm1}
\end{equation}

\end{widetext}
Using the moments  Eqs.
(\ref{eqAVN},
\ref{eqnnm1})
we get an exact expression for Mandel's parameter
$ Q = (  \langle N^2 \rangle - \langle N \rangle^2)/ \langle N \rangle -1$,
which classifies deviations from Poissonian behavior.
We now investigate limiting cases, which explain the rich Physical behaviors
of Eqs.  
(\ref{eqMR0}-
\ref{eqnnm1}).

 In the strong field limit  $\Omega>>1$
$$ P_0 \sim e^{ - {T \over 2}} \cos^2 \left( { \Omega T \over 2 } \right), $$
$$ P_1 \sim { e^{ - { T \over 2 }} \over 8 } \left[ 4 + 2 T - \left( 4 + T \right) \cos\left( \Omega T \right) \right],$$
\begin{equation}
P_2 \sim {e^{-{T\over 2}} \over 64} T \left[ (8 +  T) \cos(\Omega T) +
4 T + 16  \right].
\label{eqp106}
\end{equation}
Hence for fixed $T$,
$P_N$ shows  a bounded  oscillatory behavior, related to the 
Rabi oscillations of the excited state population.
Note that $P_1$ and $P_2$ are bounded from above according to
$$ P_1 \le 
{(8 + 3  T) e^{ - {T \over 2}} \over 8},$$ 
\begin{equation}
P_2 \le {e^{ - {T\over 2}} T \over 64}  \left( 5 T + 2 4\right).
\label{eqp2}
\end{equation}
While for $P_1$ we may set $T \to  0$ and get $P_1=1$ i.e. a $\pi$ pulse
soon to be discussed, the behavior of $P_2$ is more interesting and 
we find that we cannot reach the ideal limit of emitting two photons with
probability one. Instead we find that for a $2 \pi$ pulse $\Omega T = 2 \pi n$,
with $n$ a positive integer, the maximum of $P_2$ 
is $P_2 ^{\rm Max} \simeq 0.41$
which is found when $T = 2 ( \sqrt{61} - 1)/5$. Thus unlike the
maximum of $P_1$ which is found for a very short pulse, the maximum of
$P_2$ is found for a particular finite interaction time, which
is needed for the production of a pair of photons. This maximum is for
a fixed large Rabi frequency. The global maximum of $P_2$ is found
at $\Omega \simeq 1.25,T\simeq 4.86$ and then
$P_2 ^{\rm Max} =0.56$, namely the global maximum of $P_2$ is found
for intermediate Rabi frequencies. 

 In the mathematical limit of strong and short pulses, the product
$\Omega T$ remaining fixed,  we
have 
\begin{equation}
 \lim_{\Omega \to \infty, T \to 0} P_1 = \sin^2 \left( { \Omega T \over 2}  \right),
\label{eq106}
\end{equation}
$P_0 = 1 - P_1$, namely in this limit either one or zero
photons are emitted per pulse. The well known $\pi$ pulse
$\Omega T = n \pi$ yields the global maximum of 
$P_1$ which is unity. 

In experiments fields are never
of infinite strength and interaction time is not zero
and then exact results are useful. 
A transcendental equation for the value of $T$ which yields 
the extremum of $P_1$ is found using 
Eq. 
(\ref{eqp106})
\begin{equation}
- 2 T + ( 2 + T) \cos( \Omega T ) + 2 ( 4 + T ) \Omega \sin(\Omega T) = 0.
\label{eqp102a}
\end{equation}
Since $P_1$ is oscillating in this limit of strong fields,
there exists an infinite number of solutions of this equation.
For large $\Omega$ the maximum of $P_1$, $P_1 ^{\rm Max}$
is the first root of
this equation. For $\Omega \to \infty$, Eq. (\ref{eqp102a}) 
gives $\Omega \sin(\Omega T) = 0$  hence
the value $T\simeq n \pi/ \Omega$ yields the maximum  of $P_1$ and
\begin{equation}
P_1 ^{\rm Max} \simeq e^{ - \pi/ (2 \Omega) } \left( 1 + { 3 \pi \over 8 \Omega} \right).
\label{eqmax}
\end{equation}
This equation gives the corrections to the ideal $\pi$ pulse
limit of $\Omega \to \infty$ and $T \to 0$ which yields
$P_1 ^{\rm Max} = 1$.

The intermediate Rabi frequency  marks the transition between over damped
to under-damped behaviors. For 
$\Omega=1/2$ we have  
\begin{equation}
P_1 = { T^2 \left( 480 + 160 T + 20 T^2 + T^3 \right) \over 7680 } e^{ - {T \over 2 } }
\label{eqp105}
\end{equation}
We find $P_1 ^{\rm Max}\simeq 0.56$  when
$T\simeq 6.75$.

 The weak field
limit where $\Omega \to 0$ and $T \to
\infty$ in such a way that $\Omega^2 T$ 
is called the semi-classical  
since 
\begin{equation}
\lim_{\Omega \to 0, T\to \infty} P_N = { (\Omega^2 T)^N \over N!} e^{ - \Omega^2 T },
\label{eqPos}
\end{equation}
and  photon statistics is Poissonian. 
The semi classical maximum of $P_1$ is found when 
$\Omega^2 T = 1$ and 
%
$\lim_{\Omega \to 0, T \to \infty} P_1 ^{\rm Max}= e^{-1}$.
%
While for $P_2$ the maximum is found when  $\Omega^2 T =2$ and
then $P_2 ^{\rm Max}= 2 e^{-2}$.

\begin{figure}
\begin{center}
\epsfxsize=80mm
\epsfbox{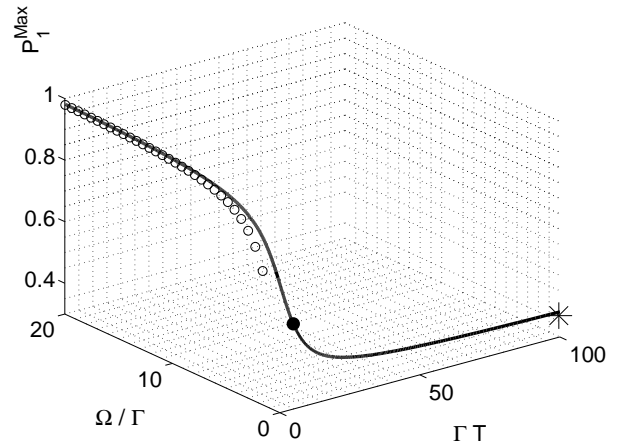}
\end{center}
\caption{The maximum of the probability of
emission of a single photon in a square
pulse
obtained from Eq. (\ref{eqMR1}) 
with  $\Gamma = 20 \mbox{MHz}$.
The circles show the strong field approximation
Eq. (\protect{\ref{eqmax}}).
The star gives the asymptotic semi-classical weak Rabi frequency  behavior
$\lim_{\Omega \to 0, T \to \infty} P_1 ^{\rm Max}= e^{-1}$.
The dark dot is for the intermediate Rabi frequency
$\Omega=1/2$ as predicted based on Eq.
\protect{(\ref{eqp105})}.
}
\label{fig9}
\end{figure}

\begin{figure}
\begin{center}
\epsfxsize=80mm
\epsfbox{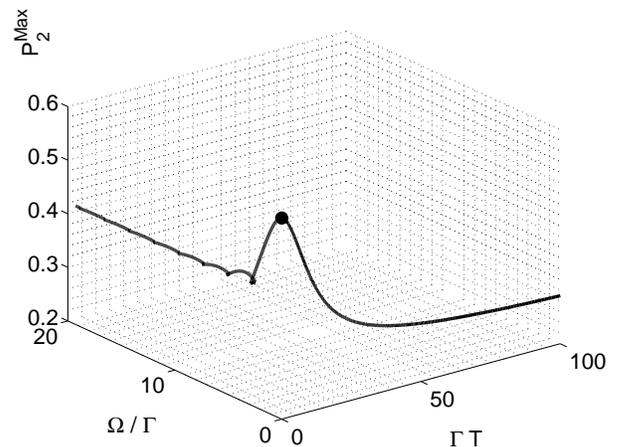}
\end{center}
\caption{ The maximum of $P_2$.   
 For weak fields the semi-classical maximum
of $P_2$ is $P_2 ^{\rm Max} = 2 e^{-2} \simeq 0.27$ while in the strong
field limit the maximum is $P_2 ^{\rm Max} \simeq 0.41$ when
$\Gamma T=2 (\sqrt{61}-1)/5$, 
as predicted in the text.
The global maximum (i.e. the dark dot) 
is found for an intermediate Rabi frequency. 
}
\label{fig20}
\end{figure}

 In Fig. \ref{fig9} we show the maximum of $P_1$ obtained using
Eq. (\ref{eqMR1}). The Fig. shows  for each Rabi frequency 
the value of the maximum of $P_1$ and the duration of pulse 
interval on which this maximum is found. The general trend
is that for strong excitation fields $P_1 ^{\rm Max}=1$ for a $\pi$ pulse, 
and
then  $P_1 ^{\rm Max}$ is decreasing as we  decrease $\Omega$,
and at the same time the value of $T$ on which the maximum is
found is increasing. A transition between a strong field and
semi-classical limit is easy to observe. 
Fig. \ref{fig20} shows the maximum of $P_2$. The three
regimes of strong fields, intermediate, and semi-classical limit
are clearly observed.  
The observation  that a peak in 
$P_2$ is found
for intermediate Rabi frequency,  makes this case
more interesting, and similar behaviors are expected also for
other types of pulse shapes. 

 For two reasons it is interesting to  investigate the probability of
emission of two photons using a short pulse. First since 
ideal  single photon emission
demands $P_2=0$ and secondly since inter-arrival times of
a pair of photons generated from a short pulse
are expected to be shorter if compared with inter-arrival times
of a pair of photons generated by a longer pulse, 
and hence  potentially  
useful for sources of indistinguishable  photon pairs. 
 In 
the limit $T \to 0$ we find
\begin{equation}
P_2 \sim {\Omega^4 T^5 \over 480}.
\label{eqT5}
\end{equation}
The $T^5$ behavior in Eq. (\ref{eqT5}) can be explained, by noting
that the field dependence of $P_2$ is expected, namely
$P_0 \propto \Omega^0, P_1 \propto \Omega^2$
and $P_2 \propto \Omega^4$. Physically  we obviously
expect that $P_2 \propto \Omega^4 T^m$ and
$m$ is a positive integer. 
As mentioned in Eq. 
(\ref{eq106})
 for a short and strong  pulse we have $P_2=0$ 
and hence me must have $m>4$, 
otherwise the product
$\Omega^4 T^m$ does not approach zero in the limit $\Omega \to \infty$ and
$T \to 0$ with $\Omega T$ remaining fixed.   And indeed we find $m=5$
in Eq. (\ref{eqT5}).

 Mandel's $Q$ parameter 
 characterizing deviation from Poissonian
photon statistics, is analyzed using Eqs.
(\ref{eqAVN},
\ref{eqnnm1}) in the limits considered so far for
the probabilities $P_N$. In the limit of long interaction times
\begin{equation}
\lim_{T \to \infty} Q = - { 6 \Omega^2 \over \left( 1 + 2 \Omega^2\right)^2}.
\label{eqMalta}
\end{equation}
This result was obtained by Mandel \cite{Mandel}
for an atomic transition
interacting with a cw laser. As expected if the interaction
time $T$ is long, 
namely many photons are emitted, the familiar
resonance fluorescence sub-Poissonian behavior is found.

 New behavior is found for finite interaction time  $T$.
For strong laser fields, $\Omega \to \infty$ we find 
using Eqs. 
(\ref{eqAVN},
\ref{eqnnm1})
%
\begin{equation}
Q \sim { - \left[ 1 - \cos \left( \Omega T \right) e^{-{3 T \over 4}}\right]^2 +
3 T \cos \left( \Omega T\right) e^{ - {3 T \over 4}} \over
2 \left[ 1 + T - \cos \left( \Omega T \right) e^{- {3 T \over 4}} \right]}. 
\label{eqQapp}
\end{equation}
When the square pulse is short and strong, 
however $\Omega T$ remains finite,
we find an interesting discontinuous behavior
\begin{equation}
\lim_{\Omega \to \infty, T \to 0} Q = \left\{
\begin{array}{l l}
 -\sin^2 \left( { \Omega T \over 2 } \right)   & \Omega T \ne 2 \pi n \\
6/7 & \Omega T = 2 \pi n,
\end{array}
\right.
\label{eqDIS}
\end{equation}
where $n$ is a positive integer. 
Of-course for finite
though large (small) values of $\Omega$ $(T)$,
$Q$ is a continuous   
function. Though for large $\Omega$,
 $Q$  exhibits: $(i)$ very sharp, cusp resonances on $\Omega T = 2 \pi n $
which become narrower as the mathematical limit of $\Omega \to \infty$
and $T \to 0$ is approached and $(ii)$ 
the value of $Q$ on the peaks is roughly $6/7$.
To understand the
origin of the  behavior predicted in Eq. 
(\ref{eqDIS}) note that 
according to 
Eq.
(\ref{eq106})
we either obtain zero photons or one photon in this limit.  
Hence we expect naively $Q = -P_1= - \sin^2 (\Omega T /2)$.
However this approach cannot be used when $P_1 \simeq 0$,
namely when $\Omega T \simeq  2 \pi n $, when
$\langle N \rangle \simeq 0$ and the fluctuations become non-trivial. 
 Instead
one must consider the limit with care and include the $1/\Omega$
corrections, which yield
the $Q\sim  6/7$ law in Eq. 
(\ref{eqDIS}).

We now consider the rapid adiabatic following method. 
In Table 1 we show photon statistics obtained 
using the generating function formalism  
Eq. (\ref{eqA01})
and experiments of
the group of Orrit (numbers in the brackets in Table 1). 
A good agreement is found, indicating that
the simple two level system approximation of the molecule,
is an excellent approximation, a conclusion reached previously
in \cite{Brunel}. 
The question remains what is the ideal choice
of control parameters for the generation of
a single photon? And how does this method compare with
the square pulse approach?  

From the experimental data in Table 1 we see that maximum
of $P_1$ for $\Omega=3.2 \Gamma$ is $P_1 = 0.68$ found for
$\Delta_{\rm RF} = 88 \Gamma$.
We found numerically (data not shown)  that for 
$\Omega=3.2 \Gamma, \nu_{\rm RF} = 3 MHz$
the maximum of $P_1$ is $P_1 \simeq 0.69$
for $\Delta_{\rm RF} \simeq 100 \Gamma$. 
We see that the experiments of the group of Orrit
are conducted very close to optimal
condition for a particular Rabi frequency, in the sense that $P_1$ is
close to its maximum.
Note that {\em in principle} for very strong Rabi frequency
and very large $\Delta_{\rm RF}$ we may obtain effectively a $\pi$
pulse, i.e. a single photon with probability one.

\begin{tabular}{lcccccr}
\hline\hline
$ \Delta_{\rm RF}$ &  $\nu_{RF}$ & $ \Omega $ & $P_0$ & $ P_1$ & $P_2 $  \\ \hline
$ 50\Gamma$& 3 \mbox{MHz} & $3.2\Gamma$ & 0.01(0.02) & 0.53(0.56) & 0.32(0.31)  \\
$ 88\Gamma$& 3 \mbox{MHz} & $3.2\Gamma$ & 0.09(0.11) & 0.68(0.68) & 0.20(0.18)  \\
$130\Gamma$& 3 \mbox{MHz} & $3.2\Gamma$ & 0.24(0.22) & 0.66(0.66) & 0.10(0.10)  \\
$160\Gamma$& 3 \mbox{MHz}  & $5.0\Gamma$ & 0.04 & 0.72(0.74) & 0.21  \\
\end{tabular} \\
{\bf Table 1:} Rapid adiabatic following method, 
theory versus experiment, $\Gamma=20  \mbox{MHz}$.

A more detailed investigation of the rapid adiabatic method
will be published elsewhere. We mention that the maximum of $P_1$ for
the square pulse is slightly larger
compared with the  rapid adiabatic method. 
More specifically we searched for the maximum
of $P_1$ in the range $0<\Omega<20 \Gamma$, 
$0<\Delta_{\rm RF} < 600 \Gamma$ 
and $\nu_{\rm RF}=3 \mbox{MHz}$  to be reasonably close to experimental
situations (see Table 1), and found $P_1 ^{\rm Max} \simeq 0.82$. This maximum
should be compared with
with $P_1 ^{\rm Max} \simeq 1$ in Fig. \ref{fig9} for the square pulse 
and for a similar range of Rabi frequency. 
We found
that the rapid adiabatic following
method has one possible advantage over the square
pulse, 
and that is the absence of fast Rabi oscillations in $P_1$,
in the limit of strong fields. If $P_1$ is oscillating
the generation of a single photon is less stable, and in this
sense the rapid adiabatic following method is better. The absence
of Rabi oscillations is due to
the smooth time dependent detuning which averages out the Rabi oscillations.

Finally, while our two level system approximation
works very well for certain low temperature single molecules \cite{Brunel}
and is expected to work well for simple atoms \cite{Darquie}, 
extensions of this work for 
multi level emitters, where even the $\pi$ pulse condition for single
photon emission is not valid 
\cite{Lounis} 
should be investigated theoretically, 
though then analytical theory is
cumbersome. 
While our work provides the conditions for the maximum of
the emission of a pair of photons, further work 
on the inter arrival times
between such pairs and their entanglement is also
of interest.  

{\bf Acknowledgment} 
This work was supported  by the National Science Foundation 
award CHE-0344930. 
EB thanks  Israel's Science Foundation
for financial
support.

\end{document}